\documentclass[reprint,
 amsmath,amssymb,
aps, 
pra,
]{revtex4-2}

\setcitestyle{super}

\usepackage{graphicx}
\usepackage{dcolumn}
\usepackage{bm}

\usepackage[utf8]{inputenc}
\usepackage[T1]{fontenc}
\usepackage{mathptmx}
\usepackage{etoolbox}
\usepackage{color}

\begin{document}


%
%
%
\title[]{Thermal analysis of GaN-based photonic membranes for optoelectronics}

\author{Wilken Seemann}
\affiliation{Institut für Festk\"orperphysik, Universit\"at Bremen, Otto-Hahn-Allee 1, 28359 Bremen, Germany}
\altaffiliation[Wilken Seemann also at:]{\,MAPEX Center for Materials and Processes, Universität Bremen, Bibliotheksstraße 1, 28359 Bremen, Germany}

\author{Mahmoud Elhajhasan}
\author{Julian Themann}
\author{Katharina\,\,Dudde}
\author{Guillaume W\"ursch}
\author{Jana Lierath}
\author{Gordon Callsen}
\email{gcallsen@uni-bremen.de}

\affiliation{Institut für Festk\"orperphysik, Universit\"at Bremen, Otto-Hahn-Allee 1, 28359 Bremen, Germany}

\author{Joachim Ciers}
\author{\AA sa Haglund}

\affiliation{Department of Microtechnology and Nanoscience, Chalmers University of Technology, 41296 Gothenburg, Sweden}

\author{Nakib H. Protik}

\affiliation{Department of Physics and CSMB, Humboldt-Universit\"at zu Berlin, 12489 Berlin, Germany}

\author{Giuseppe Romano}

\affiliation{MIT-IBM Watson AI Lab, IBM Research, Cambridge, MA, USA}

\author{Rapha\"el Butt\'{e}}
\author{Jean-François Carlin}
\author{Nicolas Grandjean}

\affiliation{Institute of Physics, \'{E}cole Polytechnique F\'{e}d\'{e}rale de Lausanne (EPFL), CH-1015 Lausanne, Switzerland}

\date{\today}

\begin{abstract}

Semiconductor membranes find their widespread use in various research fields targeting medical, biological, environmental, and optical applications. Often such membranes derive their functionality from an inherent nanopatterning, which renders the determination of their, e.g., optical, electronic, mechanical, and thermal properties a challenging task. In this work we demonstrate the non-invasive, all-optical thermal characterization of around 800-nm-thick and 150-$\mu$m-wide membranes that consist of wurtzite GaN and a stack of In$_{0.15}$Ga$_{0.85}$N quantum wells as a built-in light source. Due to their application in photonics, e.g., for vertical-cavity surface-emitting lasers, such photonic membranes are bright light emitters, which challenges their non-invasive thermal characterization by only optical means. As a solution, we combine two-laser Raman thermometry with (time-resolved) photoluminescence measurements to extract the in-plane (i.e., $c$-plane) thermal conductivity $\kappa_{\text{in-plane}}$ of our membranes. Based on this approach, we can disentangle the entire laser-induced power balance during our thermal analysis, meaning that all fractions of reflected, scattered, transmitted, and reemitted light are considered. As a result of our thermal imaging via Raman spectroscopy, we obtain $\kappa_{\text{in-plane}}\,=\,165^{+16}_{-14}\,$Wm$^{-1}$K$^{-1}$ for our best membrane, which compares well to our simulations yielding $\kappa_{\text{in-plane}}\,=\,177\,$Wm$^{-1}$K$^{-1}$ based on an \textit{ab initio} solution of the linearized phonon Boltzmann transport equation. Furthermore, we study the thermal impact of roughening of our membranes and the addition of extra semiconductor layers. Thanks to our experimental approach, all variations of $\kappa_{\text{in-plane}}$ become directly accessible to the experimentalist via highly spatially resolved temperature maps. Our work presents a promising pathway towards thermal imaging at cryogenic temperatures, e.g., when aiming to elucidate experimentally different phonon transport regimes via the recording of non-Fourier temperature distributions.

\end{abstract}

\maketitle

\section{\label{sec:Introduction}Introduction}

The field of semiconductor membranes (SMs) is constantly fueled by emerging membrane materials that are promising for a wide range of pioneering applications. The most established membrane materials are, e.g., amorphous SiN \cite{wang_fabrication_2018,shin_fabrication_2019} or crystalline Si.\cite{Graczykowski2017,Chavez-Angel2014} However, recently, also SiC,\cite{mohd_nasir_fabrication_2012} diamond,\cite{gao_electrically_2016} and Ge \cite{hanus_wafer-scale_2023} became routinely available as membrane material. In addition, many other semiconductor materials like, e.g., MoS$_2$,\cite{hirunpinyopas_potential_2020} BN,\cite{lee_functionalized_2022} group-III arsenides,\cite{lee_fabrication_1987,ding_tuning_2010,descamps_semiconductor_2023} and group-III nitrides \cite{elhajhasan_optical_2023,ciers_nanomechanical_2024} have already been used as free-standing membranes, sparking research related to a wide range of future applications. In this work we classify SMs that are inherently nanostructured (e.g., via the processing of hole lattices) as being of type I, which can also comprise surface engineering \cite{neogi_tuning_2015} to achieve the required physical properties. However, SMs can also comprise additional nanostructures like quantum wells (QWs) and quantum dots (QDs), which are ultimately key to the target application of such SMs classified as being of type II.

For type I SMs it is often the permeability for liquids or gases that motivates their usage, e.g., spanning from particle nano-filtration,\cite{lee_functionalized_2022} over electrochemical separation,\cite{gao_electrically_2016} to ionic sieving.\cite{hirunpinyopas_potential_2020} Evidently, just to name a few, SMs are not only promising for environmental applications like wastewater treatment and water desalination,\cite{lee_functionalized_2022} but even for medical applications like hemodialysis.\cite{gao_electrically_2016} Often the permeability of SMs is accompanied by a certain transmittance for x-rays and electron beams, which renders SMs materials like SiN and SiC ideal as sample holders for various related microscopic techniques.\cite{fu_low_2019} In addition to such almost trivial microscopy applications, further complex nanopatterning of semiconductor membranes can enable the fabrication of x-ray optics.\cite{beijersbergen_silicon_2004,vila-comamala_advanced_2009,rosner_exploiting_2018}

In the field of optics, one can already find SMs of type II with potential applications in quantum communication and computing. Here, the first pioneering structures were based on group-III arsenide membranes comprising InGaAs QDs as light emitters.\cite{ding_tuning_2010} Similar QD structures are also in use for membrane-based quantum transport devices.\cite{descamps_semiconductor_2023} Often a certain tuneability of the electronic structure of such QDs by, e.g., external fields is the key ingredient to the desired functionality. Thus, tuning of the optical and/or electronic properties of QDs by strain fields \cite{ding_tuning_2010} or electric fields \cite{descamps_semiconductor_2023} is significantly simplified by the application of SMs due to their reduced thicknesses, which boost the impact of the external fields. In addition to their use in the quantum realm, SMs of type II can also be key to macroscopic devices as detailed in the following.

For the realization of flexible electronics it is often necessary to transfer SMs of type II onto flexible substrates,\cite{chang_semiconductor_2024} because any direct material deposition is hindered by a low chemical and temperature robustness of such substrates. As a result, the main functionality of the final device arises from SMs that can, e.g., build on well-established Si-based electronics. In a figurative sense, similar challenges are faced in the field of group-III nitride based photonics, aiming, e.g., for the realization of ultraviolet (UV) light-emitting diodes (LEDs) and laser diodes (LDs). Especially for the latter case, vertical-cavity surface-emitting lasers (VCSELs) based on group-III nitride SMs of type II were demonstrated for the blue \cite{holder_demonstration_2012,holder_nonpolar_2014} and UV \cite{hjort_310_2021,cardinali_low-threshold_2022,torres_ultraviolet-b_2024} spectral range. Here, especially the ongoing development from UV-A (320\,-\,400\,nm), over UV-B, \cite{torres_ultraviolet-b_2024} to even UV-C ($\leq\,280\,\text{nm}$) VCSELs \cite{zheng_algan-based_2021} poses strong prerequisites for the material deposition by, e.g., metal-organic vapor phase epitaxy (MOVPE). The deposition of monolithic structures that comprise, e.g., distributed Bragg reflectors (DBRs) \textit{and} the optically active material [e.g., (Al)GaN/AlGaN QW stacks] proved increasingly challenging the shorter the emission wavelength. Here, the in-plane lattice mismatch and the achievable AlGaN compositions challenge the development of deep UV DBRs with reflectivities $>\,99\%$.\cite{hjort_310_2021} Thus, the required bottom DBR structure can, e.g., also be fabricated by reactive radio frequency (RF) magnetron sputtering, which is followed by a transfer of type II SMs grown by MOVPE,\cite{bergmann_electrochemical_2019} containing a stack of QWs as the UV light source. Subsequently, the top DBR structure can, e.g., be realized by sputtering in a separate process. This sequentialization of the VCSEL fabrication process based on different growth methods boosted the interest in, so-called, \textit{photonic} SMs of type II that comprise nanostructures like QWs or QDs as built-in light sources.

Consequently, the analysis of the optical, electronic, mechanical, and thermal properties of such photonic SMs of type II constitutes a challenge for joint material characterization efforts. Thus in this work, driven by optical applications related to, e.g., VCSELs, we demonstrate an interlinked optical and thermal analysis of photonic SMs made from group-III nitrides. We combine two-laser Raman thermometry (2LRT) with (time-resolved) photoluminescence (TRPL) spectroscopy to obtain the in-plane thermal conductivity $\kappa_{\text{in-plane}}$ of SMs with different backside roughnesses and layer sequences. Such controllable backside roughening is linked to our state-of-the-art SM fabrication process based on electrochemical etching of a sacrificial layer made by, e.g., highly $n$-type doped GaN.\cite{ciers_smooth_2021} Understanding the optical and thermal impacts of a controllable surface roughening is decisive for improving the performance of VCSELs under high injection conditions. In addition, the required thermal characterization must be non-invasive and compatible with the size of the photonic SMs, which would commonly motivate the application of standard Raman thermometry, which we call one-laser Raman thermometry (1LRT). However, recently, Elhajhasan \textit{et\,al.} \cite{elhajhasan_optical_2023} have demonstrated that 1LRT is often not suitable for measuring $\kappa_{\text{in-plane}}$ of SMs made from group-III nitrides. With 1LRT a single laser is acting as the heating laser, which simultaneously provides the temperature probing via the (resonant) Raman signal. Thus, the laser-induced heating is only monitored at the heat spot, which proves to be a main drawback of 1LRT measurements. Even though the absorbed laser power can be obtained for the analysis of 1LRT data, it remains questionable over which \textit{volume} the heating occurs. Simple measurements of the laser focus spot size are not sufficient, because the mean free paths (mfps) of phonons $l_{\text{mfp}}$ that significantly contribute to thermal transport - known as "thermal phonons" - must be considered. However, the impact of thermal phonons exhibiting certain $l_{\text{mfp}}$ values on the thermal conductivity $\kappa$ is often a hidden experimental observable that scales with the overall structural quality of the sample, doping concentrations, interface densities and roughnesses, as well as further inherent nanostructuring given by, e.g., a hole lattice among other factors. Interestingly, 2LRT measurements allow bypassing some of these fundamental challenges, which is also accompanied by other experimental advantages over 1LRT as highlighted in this work. Finally, our thermal imaging by 2LRT yields temperature distributions that render differences in $\kappa_{\text{in-plane}}$ easily accessible to the experimentalist.

Hereafter, the paper is structured as follows: In Sec.\,\ref{sec:Experiment} we first introduce our SM samples (Sec.\,\ref{subsec:Sample}) as well as our experimental techniques, challenges, and solutions (Sec.\,\ref{subsec:Experiment}) that will form the basis for the envisaged thermal imaging. Subsequently in Sec.\,\ref{sec:Results}, we demonstrate this thermal imaging based on several 2LRT mapscans. First, we show a large-scale 2LRT mapscan in Sec.\,\ref{subsec:HRmaps}, which allows us to demonstrate the extraction of $\kappa$ for different in-plane directions of our SMs. Here, a special focus rests on a reliable determination of the absorbed laser power that leads to a local heating of our samples during 2LRT measurements. Therefore, we demonstrate how we can derive a good approximation of the laser-induced heating power based on combining our 2LRT measurements with time-resolved photoluminescence analyses in Sec.\,\ref{subsec:Pabs}. Second, after having established the experimental basis of our all-optical thermometry, we proceed with analyzing the impact of backside roughness variations on $\kappa$ in Sec.\,\ref{subsec:roughness}, which is followed by a comparison to our \textit{ab initio} modeling results in Sec.\,\ref{subsec:modeling}. Hereafter, based on a detailed comparison of our experimental and theoretical results, we discuss the role of the phonon mean free paths for thermal transport in GaN in Sec.\,\ref{subsec:mfp}. Finally, our conclusions are given in Sec.\,\ref{sec:Conclusions}.
\section{\label{sec:Experiment}Experimental section}
\begin{figure*}[]
    \includegraphics[width=\linewidth]{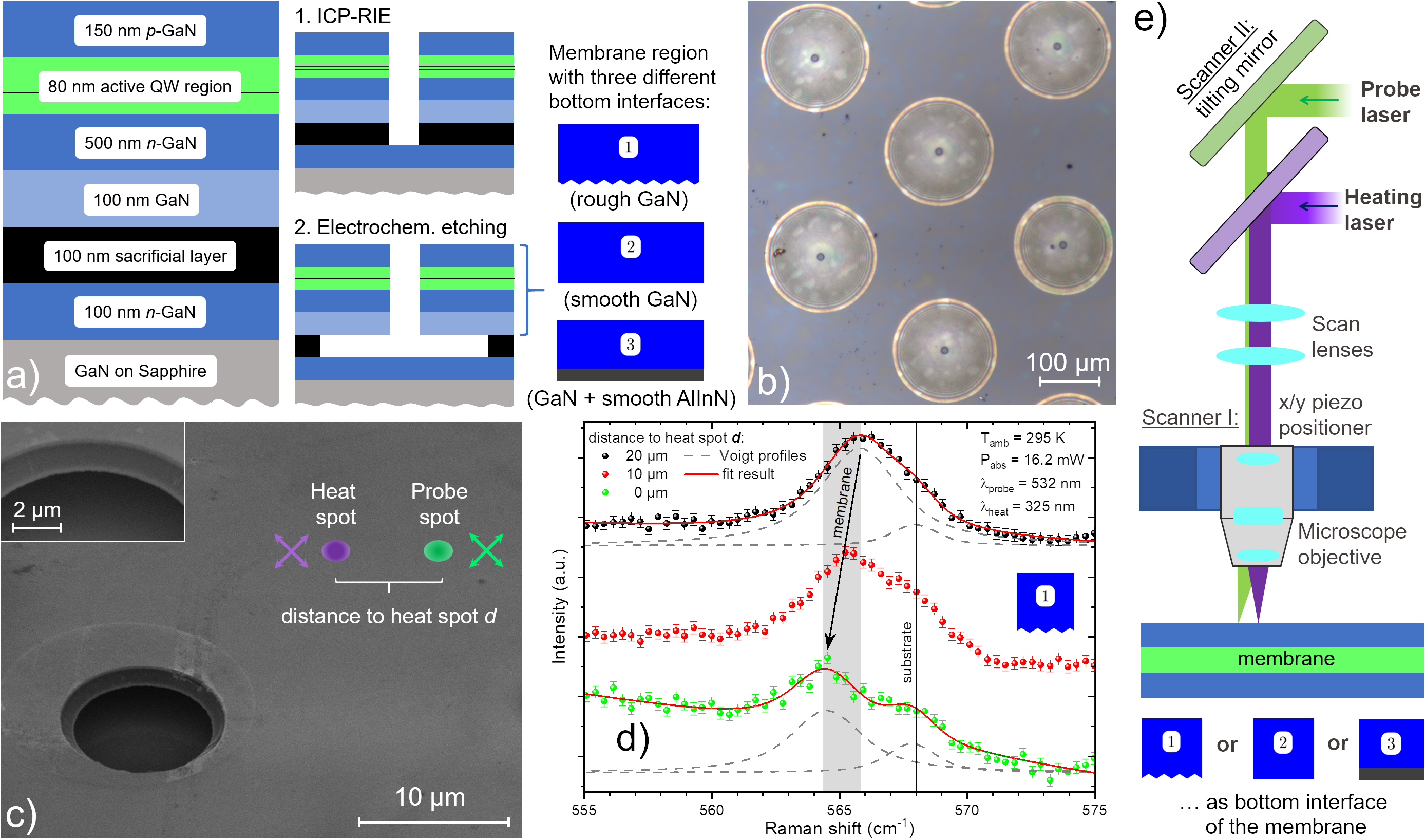}
    \caption{Experimental base: (a) Layer sequence of our group-III nitride samples grown on sapphire, which enable the fabrication of a photonic membrane by removal of the sacrificial layer. First, $\approx\,$10-$\mu$m-wide holes are fabricated by ICP-RIE, before the sacrificial layer made from highly \textit{n}-type doped GaN is removed by electrochemical wet-etching. As a result, we obtain three samples (1\,-\,3) that either exhibit a rough or smooth GaN (Al$_{0.82}$In$_{0.18}$N) backside. (b) Top-view light microscope image of a photonic membrane (sample 1) that mainly consists of \textit{c}-plane GaN. (c) Side-view scanning electron microscopy (SEM) image of sample 1, showing the etched hole and the membrane surface. The inset illustrates the smooth sidewall of the GaN membrane. The purple and green spots sketch the two laser spots that can be scanned over the membrane surface for our thermometry measurements based on Raman spectroscopy. (d) Raman spectra showing the $E_{2}^{\text{high}}$ modes of the GaN membrane and the underlying GaN layer separated by an airgap for different distances between the heating ($\lambda_{\text{heat}}=325\,\text{nm}$) and the probe ($\lambda_{\text{probe}}=532\,\text{nm}$) laser spots. (e) Sketch of the 2LRT setup that enables the movement of the laser spots over the sample surface. As a result, one can measure temperature maps that enable the determination of the in-plane thermal conductivity $\kappa_{\text{in-plane}}$.}
    \label{fig:sample}
\end{figure*}
\subsection{\label{subsec:Sample}Sample details}

Figure\,\ref{fig:sample}(a) introduces the group-III nitride layer stack grown by MOVPE in an Aixtron 200/4 RF-S horizontal reactor.  First, a low-temperature, wurtzite GaN buffer (\textit{c}-plane) is grown on a \textit{c}-plane sapphire substrate, which is followed by a 100-nm-thick layer of \textit{n}-type GaN. The latter layer is doped with Si with a concentration of $n_{\text{Si}}\,=\,3\,\times\,10^{18}\,\text{cm}^{-3}$. Subsequently, an $\approx\,100\text{-nm-thick}$, highly \textit{n}-type sacrificial layer is deposited ($n_{\text{Si}}\,=\,1\,\times\,10^{19}\,\text{cm}^{-3}\,-\,1\,\times\,10^{20}\,\text{cm}^{-3}$), which is the key part for the formation of SMs by a selective, electrochemical underetching process described in the following paragraphs. The sample growth continues with 100\,nm of non-intentionally doped (nid) GaN, $500\,\text{nm}$ of $n$-type GaN ($n_{\text{Si}}\,=\,3\,\times\,10^{18}\,\text{cm}^{-3}$), an 80-nm-thick multi-QW region, and an $\approx 150$-nm-thick top-layer mostly comprising \textit{p}-type GaN ($n_{\text{Mg}}\,=\,1\,-\,5\,\times\,10^{19}\,\text{cm}^{-3}$), which terminates this LED-like structure. More details can be found in Ref.\,\onlinecite{ciers_smooth_2021}. The multi-QW region comprises three 2.8-nm-thick In$_{0.15}$Ga$_{0.85}$N QWs separated by 15-nm-thick GaN barrier layers with $n_{\text{Si}}\,=\,1\,\times\,10^{18}\,\text{cm}^{-3}$. This selection of the QW thickness and composition ensures comparably bright light emission in case of embedment in a full LED or even LD structure.

The transformation of the monolithic material stack into a photonic SM starts with photolithography and inductively coupled plasma - reactive ion etching (ICP-RIE) of $\approx$\,10-$\mu$m-wide holes, which will allow an etchant to attack the sacrificial layer, cf. Fig.\,\ref{fig:sample}(a). Subsequently, electrochemical etching is undertaken in a suitable electrolyte (HNO$_{3}$, 0.3\,M), while the top sample surface is protected by a photoresist. Therefore, Al contacts need to be processed on our structures to apply the required bias voltage between the sample and a graphite electrode in the electrolyte bath. Further experimental details regarding this electrochemical etching process can be found in literature.\cite{bergmann_electrochemical_2019,ciers_smooth_2021} Interestingly, this etching process can be controlled by the applied voltage, the free carrier concentration in the sacrificial layer, and its overall layer sequence, which can promote internal polarization fields. As detailed by Ciers \textit{et\,al.},\cite{ciers_smooth_2021} either no etching, porosification, or a complete removal of the sacrificial layer can be achieved. In Fig.\,\ref{fig:sample}(a) we introduce three samples (1\,-\,3), which only differ by the composition of the sacrificial layer and/or the applied processing voltages. Thus, we expect that samples 1\,-\,3 should be identical regarding their emission properties as the multi-QW region should remain unaffected by the etching process. This expectation will be brought to a test in Sec.\,\ref{subsec:Pabs}.

For samples 1 and 2 the sacrificial layer is identical and consists of a 100-nm-thick layer that comprises a super-lattice (SL) of doped ($n_{\text{Si}}\,=\,1\,\times\,10^{20}\,\text{cm}^{-3}$) and nid GaN layers with a thickness of $2\,\text{nm}$ each. We found that the realization of such an SL sacrificial layer allows to maintain an improved surface morphology during MOVPE growth.\cite{ciers_smooth_2021} Here, the only difference between samples 1 and 2 is the applied processing bias voltage of either 10\,V or 6\,V, respectively. As a result of the increased voltage, sample 1 exhibits a rougher backside surface compared to sample 2, while the frontside roughness remains identical due to the aforementioned protection by the photoresist as monitored by atomic force microscopy (AFM). Further improvements regarding the backside surface roughness of our SMs can be achieved by modifying the composition of the sacrificial layer. Therefore, sample 3 comprises an $\approx\,120\text{-nm-thick}$ layer stack consisting of \textit{n}-type GaN ($n_{\text{Si}}\,=\,1\,\times\,10^{19}\,\text{cm}^{-3}$) that acts as the sacrificial layer and 1\,nm of AlN (nid) as well as 20\,nm of Al$_{0.82}$In$_{0.18}$N (nid) as an etch stop. Here, this particular composition of the AlInN layer was chosen to enable an almost lattice-matched growth \cite{butte_current_2007} with respect to GaN, cf. Fig.\,\ref{fig:sample}(a). Application of a processing voltage of 17\,V for sample 3 yields a superior surface morphology for its backside compared to samples 1 and 2. However, it remains to be tested how the addition of the AlN/Al$_{0.82}$In$_{0.18}$N etch stop layers will affect the thermal properties of sample 3. Further details regarding the sample backside morphologies will be given by AFM in the context of Fig.\,\ref{fig:roughness}.

Figure\,\ref{fig:sample}(b) shows an optical microscope image of sample 1, illustrating the $\approx\,150\text{-$\mu$m-wide}$ SMs, which mainly consist of GaN. Here, the optical image contrast arises from the $\approx\,100\text{-nm-wide}$ airgap below the SMs. A more detailed view of the frontside morphology of sample 1 is given in Fig.\,\ref{fig:sample}(c) based on scanning electron microscopy (SEM). Here the inset of Fig.\,\ref{fig:sample}(c) provides details regarding the smoothness of the edge of the central hole in the SM prepared by ICP-RIE. Furthermore, two laser spots are sketched on the SM shown in Fig.\,\ref{fig:sample}(c). The depicted double laser excitation will form the basis for the optical and thermal analysis of all samples by 2LRT, which is further explained in the following Sec.\,\ref{subsec:Experiment}. 

\subsection{\label{subsec:Experiment}Experimental details and challenges}

An example of typical Raman spectra that illustrate the transition from a heated to an unheated situation on the SMs is depicted in Fig.\,\ref{fig:sample}(d). The Raman selection rules are strongly obeyed for our \textit{c}-plane SMs mainly consisting of GaN, which means that for Raman measurements with our probe laser in an unpolarized, backscattering configuration z(.,.)\=z as depicted in Fig.\,\ref{fig:sample}(d), only the $E_{2}^{\text{low}}$ and $E_{2}^{\text{high}}$ modes appear as Raman-active modes. For the following Raman thermometry based on laser-induced heating, it represents an advantage to rely on such non-polar Raman modes as the $E_{2}$ modes, because no coupling to free carriers occurs.\cite{Nenstiel2015a} Such free carriers could, e.g., be induced by doping, defects, or even by the laser excitation itself. From the two main Raman modes, the $E_{2}^{\text{low}}$ is known for its particular small mode shifts under the impact of, e.g., stress \cite{callsen_phonon_2011} and temperature.\cite{bagnall_simultaneous_2017} As a result, our focus rests on the $E_{2}^{\text{high}}$ mode of GaN situated around 567\,cm$^{-1}$ in high quality GaN \cite{Haboeck2003b,Kirste2011} at an ambient temperature of $T_{\text{amb}}\,=\,295\,\text{K}$. It shall be noted that in this work we only focus on Raman thermometry based on our probe laser, while the heating laser is exclusively used for heating, but not for any additional probing of temperatures via the resonant Raman signal.

Interestingly, in Fig.\,\ref{fig:sample}(d) we can observe two $E_{2}^{\text{high}}$ modes in the corresponding energy range. The first $E_{2}^{\text{high}}$ mode at lower wavenumbers belongs to the SMs, while the second $E_{2}^{\text{high}}$ mode, situated around  568\,cm$^{-1}$, is linked to the \textit{n}-type GaN beneath the sacrificial layer, cf. Fig.\,\ref{fig:sample}(a). The wavenumber position of the second $E_{2}^{\text{high}}$ mode is typical for the growth of GaN on \textit{c}-plane sapphire.\cite{park_comparison_2015} The apparent small mode separation is a central experimental challenge for the thermometry in this paper. Thus, our mode assignment is further justified by comparing different situations of laser-induced heating that are obtained based on the experimental setup sketched in Fig.\,\ref{fig:sample}(e). Based on this setup we can control the distance $d$ between the heating and probe laser focus spots that were introduced in Fig.\,\ref{fig:sample}(c). Thus, in case of sufficient laser-induced heating, the spectral splitting between the aforementioned two $E_{2}^{\text{high}}$ modes can be increased by decreasing $d$ from 20\,$\mu$m, over 10\,$\mu$m, down to 0\,$\mu$m as shown in Fig.\,\ref{fig:sample}(d). It is evident that only the first $E_{2}^{\text{high}}$ mode shifts with decreasing values of $d$, while the second $E_{2}^{\text{high}}$ mode remains constant as shown in Fig.\,\ref{fig:sample}(e). Thus, when the laser focus is set to the top surface of the SMs, the heating power of the heating laser does not suffice to heat the bulk structure under the airgap. It shall be noted that all our 2LRT measurements take place in vacuum, rendering this a vacuum gap or just gap for the sake of simplicity. In contrast, the SMs can be heated, which is key to the optical thermometry described in this work. However, first some additional important experimental details of thermometry by 2LRT remain to be clarified in the following Secs.\,\ref{subsubsec:Raman}\,-\,3:

\subsubsection{\label{subsubsec:Raman}Raman and photoluminescence detection}

The laser-induced heating is obtained via a continuous wave (cw) UV laser ($\lambda_{\text{heat}}\,=\,325\,\text{nm}$), which is guided towards the sample via a suitable beamsplitter, a 4f-lens system (labeled "scan lenses"), and a 50x UV microscope objective with a numerical aperture (NA) of 0.40 as depicted in Fig.\,\ref{fig:sample}(e). The corresponding heating laser focus spot can be moved over the sample via a closed-loop \textit{x}/\textit{y} piezo positioner labeled "scanner I". The sample itself is placed in a vacuum chamber with a base pressure of $\approx\,1\,\times\,10^{-6}\,\text{mbar}$ to avoid any convection cooling by gases. The Raman signal required for the temperature probing is generated via a cw probe laser ($\lambda_{\text{probe}}\,=\,532\,\text{nm}$) that is guided towards the sample via a two-axes piezo-based tilting mirror ("scanner II") before following the beam path of the heating laser, cf. Fig.\,\ref{fig:sample}(e). As a result, tilting the probe laser beam via scanner II enables a scanning of the focused probe laser spot over the sample surface, while the heating laser focus spot remains fixed in a position of choice. As no tilting of the heating laser beam is introduced via scanner II, the 4f-lens system simply focuses and collimates the heating laser beam passing through its center. As a result, the heating and probe laser focus spots can independently be moved on the surface of the samples, which enables the recording of Raman mapscans under the presence of different heating laser powers. The corresponding absorbed power of the heating laser is denoted by $P_{\text{abs}}$ in Fig.\,\ref{fig:sample}(d). The corresponding power of the probe laser is always low enough to avoid any additional and unwanted heating. The detection beam path for the Raman signal related to $\lambda_{\text{probe}}$ or the photoluminescence (PL) signal related to $\lambda_{\text{heat}}$ is not illustrated in Fig.\,\ref{fig:sample}(d) for the sake of simplicity. For detection we utilized a single, triple-grating monochromator equipped with suitable filters and a charge-coupled device (CCD), providing a spectral resolution of $\leq\,1\,\text{cm}^{-1}$ around 532\,nm (1800\,l/mm grating). For overview PL spectra we employ a 150\,l/mm grating.

The appearance of two $E_{2}^{\text{high}}$ modes in Fig.\,\ref{fig:sample}(d) is directly linked to the focal depth of the microscope objective in use. We employ a, so-called, 3-wavelength microscope from Sigma Koki, which exhibits sufficient transmission from the UV to the visible spectral region. Furthermore, we experienced an excellent performance of this microscope objective regarding its ability to focus our heating and probe lasers simultaneously, while keeping both lasers in focus during the scanning action of scanners I and II. However, the limited focal depth of $1.6\,\mu\text{m}$ around 550\,nm limits the ability of separating the SMs Raman signal from the Raman signal of the underlying bulk material. Nevertheless, it is already an achievement that the signal is as clearly separable as that shown in Fig.\,\ref{fig:sample}(d). In order to ensure a high reproducibility for our measurements, we implemented a focusing routine in our setup based on a motorized focusing stage with a step resolution of 100\,nm. The smallest spot sizes for the probe and heating laser were found in case the PL signal of the QWs embedded in our structures was maximized while focusing. The precise determination of the laser spot sizes was achieved by PL and Raman mapscans with a step resolution of $\leq\,300\,\text{nm}$. Further details regarding our experimental procedure are given in Ref.\,\onlinecite{elhajhasan_optical_2023}. For our 2LRT measurements we obtain a full width at half maximum of $d_{\text{heat}}=1400\,\pm\,180\,\text{nm}$ for the focus spot diameter of our heating laser ($\lambda_{\text{heat}}=325\,\text{nm}$) and $d_{\text{probe}}=1360\,\pm\,160\,\text{nm}$ for our probe laser ($\lambda_{\text{probe}}=532\,\text{nm}$).
\\

\subsubsection{\label{subsubsec:TRPL}Time-resolved photoluminescence}

For TRPL measurements the heating laser is replaced by a pulsed laser ($\lambda_{\text{TRPL}}\,=\,266\,\text{nm}$) with a pulse duration ($t_{\text{pulse}}$) of 80\,ps and a variable repetition rate ($r_{\text{rep}}$) that is adapted to the experimentally recorded time constants ($r_{\text{rep}}\,=\,1\,\text{MHz}$). Thus, we can ensure that after the excitation of the sample with a single light pulse, it always fully returns to equilibrium. For the detection of the TRPL signal we employ a fast photomultiplier tube (transient time spread $\leq\,150\,\text{ps}$) and standard time-correlated single photon counting (TCSPC) electronics. Furthermore, any potential detrimental pile up effect, which is typical for the employed TCSPC, is avoided by adopting the photon count rates to the laser repetition rates.

\subsubsection{\label{subsubsec:2LRT}Acquisition of temperature maps}

Commonly, either the Raman mode shift a), the Raman mode broadening b), or the Raman mode intensity c) can be employed as a thermometer based on a suitable calibration. For the latter case c) it is often convenient to use the intensity ratio of a selected Raman mode on the Stokes- and anti-Stokes-side of the Raman spectrum. However, the resulting temperature resolution is often limited \cite{beechem_micro-raman_2008}, which is in our case further worsened by strong background PL that troubles the determination of Raman mode intensities. This phenomenon is especially troublesome in case of 2LRT measurements based on a heating and probe laser. The heating laser causes a strong PL signal that is, e.g., composed of GaN bandedge, QW, and defect luminescence traces distributed over a broad spectral range (further details follow in the context of Fig.\,\ref{fig:tart}). Thus, contrary to conventional Raman spectroscopy, the Stokes- \textit{and} anti-Stokes-side are often impacted by background PL, which can only be (partially) avoided by a careful selection of $\lambda_{\text{probe}}$. However, the flexibility regarding the choice of $\lambda_{\text{probe}}$ is limited by the available microscope objectives that face the challenge of focusing the heating and probe laser simultaneously. In general, the Raman mode broadening b) represents the most suited temperature gauge, because any impact of temperature-induced stress is minimized contrary to the case of the Raman mode shift a). However, the available spectral resolution often sets the limit for the general applicability of the Raman mode broadening. Thus, for limited temperature rises it often proves more practical to utilize the Raman mode shift for 2LRT measurements as discussed in our previous study \cite{elhajhasan_optical_2023} and also employed in this work.

Our procedure for measuring temperature maps by 2LRT is detailed in the following paragraphs A\,-\,C:

A) First, a micro-PL ($\mu$PL) mapscan is recorded by mapping the heating laser over the sample surface based on scanner I, cf. Fig.\,\ref{fig:sample}(e). Therefore, the heating laser is first pre-focused onto the sample surface via a home-made microscope integrated into our 2LRT setup (not shown) and subsequently fine-focused based on the QW luminescence of our sample under low injection conditions. We found that this procedure provides the best compromise for simultaneously focusing the heating and the probe lasers. The resulting $\mu$PL mapscan enables the determination of the heating laser focus spot size via the edge of the central hole in our SMs. This method resembles the common "knife-edge" method,\cite{de_araujo_measurement_2009} however, it utilizes the inherent lateral structuring of our photonic SMs. \cite{elhajhasan_optical_2023} In addition, the $\mu$PL mapscan enables the precise positioning of the heating laser's focus point with respect to the $\approx\,10$-$\mu$m-wide hole in the center of each SM.

B) Subsequently, the temperature probe laser is engaged and its focus spot on the sample surface is overlapped with the focus spot of the heating laser based on the microscope image. Recording of a conventional Raman mapscan with support of scanner II and the 4f-lens system allows us to obtain an absolute position referencing between our two scanner systems. The first Raman mapscan is recorded while the heating laser is shuttered, which allows us to map the strain distribution across our sample. We label such a Raman mapscan an "unheated" mapscan, which simultaneously allows us to determine the focus spot size of the probe laser via scans over the edge of the central hole in the SMs. 

C) A second "heated" mapscan is recorded while the heating laser is engaged. Subsequently, all Raman spectra that belong to the heated and unheated mapscans are fitted in the spectral region of interest as exemplified in Fig.\,\ref{fig:sample}(d). As fit function we use the sum of a linear function and two Voigt profiles with a fixed Gaussian broadening. This Gaussian broadening is given by the measured spectral resolution of our Raman setup. An exemplary result of our fit procedure is shown by the red solid line that fits the Raman spectrum for \textit{d}\,=\,20\,$\mu$m in Fig.\,\ref{fig:sample}(d). Here, only the two Voigt profiles are also shown by gray dashed lines, while the linear function is not sketched for the sake of clarity. After applying our fit procedure we obtain an unheated and a heated mapscan of the Raman mode positions, which can be subtracted from each other. As a result, we obtain a mapscan of the differential Raman mode shift, which is predominantly caused by the laser-induced heating. Subsequently, we apply our, so-called, local temperature calibration to translate differential Raman mode shifts into local temperatures. The subtraction of mapscans allows us to remove the minor impact of strain fluctuation across our sample surface, which improves the accuracy of our final temperature maps. At a maximum such strain fluctuations cause additional Raman mode shifts of $\approx\,\pm\,0.15\,\text{cm}^{-1}$, which is small compared to the maximal temperature-induced Raman mode shifts on the order of 1\,cm$^{-1}$ as shown in Fig.\,\ref{fig:sample}(d). 

We recorded the required local temperature calibration on sample 1 and cross-checked the result with our other samples. For the temperature calibration we recorded Raman spectra for a fixed distance to the hole in our SMs ($d\,=\,20\,\mu\text{m}$) with a Raman system almost identical to the one in use for the 2LRT measurements. The only minor difference is an even improved spectral resolution of $\approx\,0.6\,\text{cm}^{-1}$ around $\approx\,532\,\text{nm}$. Also for the temperature calibration the sample is positioned in a vacuum chamber provided by a customized heat stage (HCS421VXY) made by Instec. As a result of our local temperature calibration, we obtain the spectral position of the $E_{2}^{\text{high}}$ mode over temperature. After inverting this trend, it is used as our thermometer for all subsequent Raman thermometry experiments.


\section{\label{sec:Results}Results and discussion}

\subsection{\label{subsec:HRmaps}Temperature maps with high spatial resolution}
\begin{figure*}[]
    \includegraphics[width=\linewidth]{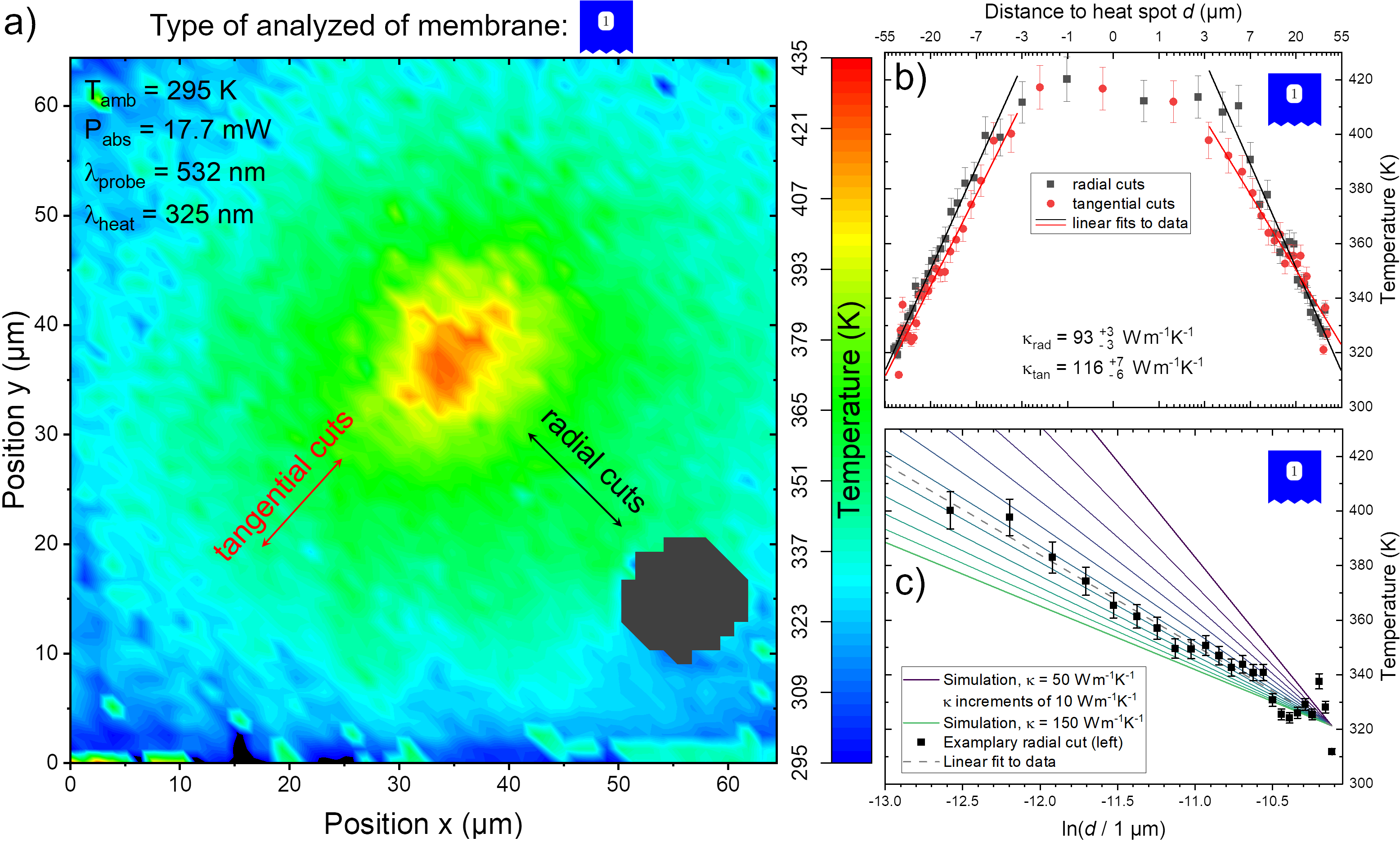}
    \caption{Thermal analysis of sample 1 by 2LRT: (a) Temperature mapscan (step resolution $\approx\,1.3\,\mu\text{m}$) obtained from the temperature-induced Raman shift. The hole in the membrane and the position of the laser-induced heat spot ($\lambda_{\text{heat}}\,=\,325\,\text{nm}$) are clearly visible. (b) Temperature profile cuts across the heat spot for radial (black) and tangential (red) directions. Please note the symmetrical double log-scale of the axis showing the distance to the heat spot $d$. For displaying reasons this axis is linearized for $\mid d \mid\,\leq\,1\,\mu\text{m}$. The solid lines display linear fits to the data, indicating an in-plane thermal anisotropy. (c) Comparison of one side of a measured radial temperature profile (symbols) with temperature gradients simulated for different $\kappa$ values (solid lines). As a result, we obtain $\kappa_{\text{in-plane}}$ and its error interval from 2LRT measurements.}
    \label{fig:2LRT}
\end{figure*}

Figure\,\ref{fig:2LRT}(a) shows the result of 2LRT at $T_{\text{amb}}=295\,\text{K}$ on sample 1 based on a temperature map over 60\,$\times$\,60\,$\mu$m$^2$. This temperature map is the result of 2\,$\times$\,2601 Raman spectra with an integration time of 10\,s each, resulting in a total measurement time of $\approx$\,15\,hours (which includes scanner displacement and settling times). Recording of such a temperature map requires a superb position stability regarding the two focus spots of the heating and the probe lasers, but also the sample itself. Otherwise, the subtraction of the heated and unheated map would yield a non-symmetric temperature map due to spatial drift. Over such extended recording times we can only observe minor total spatial drifts of $\leq\,300\,\text{nm}$ in our fully customized and therefore also very stable experimental apparatus, which is significantly smaller than the step resolution of $\approx\,1.3\,\mu\text{m}$ chosen for Fig.\,\ref{fig:2LRT}(a). Thus, a sub-pixel drift compensation did not prove necessary for the present work that only studies SM structures with rather large features (e.g., the central hole) on the scale of micrometers. However, we wish to note that the presented 2LRT result relies on a reasonable temperature stability in the laboratory ($\,\pm\,1\,\text{C}^{\circ}$) and active temperature stabilization of the main components of the setup (e.g., breadboard for all optics, microscope objective holder) to better than $\,\pm\,0.1\,^{\circ}\text{C}$.

As a result of the temperature map from Fig.\,\ref{fig:2LRT}(a) one can now extract temperature gradients for all in-plane directions of choice. Here, we decided to plot the temperature gradients for radial and tangential directions with respect to the central hole in the SM. Corresponding temperature cuts are plotted in Fig.\,\ref{fig:2LRT}(b) in a symmetrical double log-scale plot, which exhibits a linearized inner region ($\mid d \mid\,\leq\,1\,\mu\text{m}$) for displaying reasons. From a simple analytical model based on Fourier's law we can derive a basic understanding for the temperature ($T$) cuts shown in Fig.\,\ref{fig:2LRT}(b). For a thin membrane with thickness $t$ with full radial symmetry and a point-like heat source that injects the heat power $P_{\text{heat}}$, one obtains the following expression for the thermal conductivity $\kappa$:\cite{carslaw_conduction_1959,Graczykowski2017}

\begin{equation}
\label{eq:1}
\kappa = - \frac{ P_{\text{heat}} }{2 \pi\,t} \left(   \frac{\text{d}\,T     } {\text{d}\,\text{ln}(d/1\,\mu m)}   \right)^{-1}.
\end{equation}

Here, $d$ is the distance to the central heat source in micrometer and $\kappa$ is assumed as independent from $T$ for sufficiently small values of $\text{d} T$. As a result, measured temperature cuts plotted over $d$ should be linear in the logarithmic part of a double log-scale plot as shown in Fig.\,\ref{fig:2LRT}(b). Thus the slope of the plotted temperature gradients is inversely proportional to $\kappa$. Interestingly, we find significantly different slopes for the radial (black symbols) and tangential (red symbols) temperature cuts. Here the depicted error bars for $T$ originate from the fit uncertainty of the Voigt peak positions. The overall experimental situation, however, requests a more sophisticated numerical modeling, which takes the full sample geometry (SM with a central hole), the light penetration depths of the heating and probe lasers, and the corresponding beam profiles into account. Details regarding our numerical model that we implemented in \texttt{COMSOL} based on a full three-dimensional solution of the heat transport equation can be found in the Supplemental Material of Ref. \onlinecite{elhajhasan_optical_2023}. The required light penetration depths at $\lambda_{\text{heat}}$ and $\lambda_{\text{probe}}$ are taken from literature.\cite{kawashima_optical_2024}

Based on our numerical model we can derive temperature gradients for a range of $\kappa$ values as depicted in Fig.\,\ref{fig:2LRT}(c) to obtain a direct comparison to our experimental data. The comparison of a linear fit [gray dashed line in Fig.\,\ref{fig:2LRT}(c)] to the data points and the corresponding set of numerical temperature gradients enables the determination of $\kappa$ for the radial and tangential direction. Please note that the $\kappa$ increment in our numerical simulations is by a factor of four finer than that shown in Fig.\,\ref{fig:2LRT}(c) for displaying reasons. In addition, we can derive the error interval for all $\kappa$ values by comparing the slope uncertainty of the linear fit (gray, dashed line) to our modeled temperature gradients. As a result, we obtain for the radial and tangential directions indicated in  Fig.\,\ref{fig:2LRT}(a) either $\kappa_{\text{rad}}\,=\,93^{+3}_{-3}\,\text{Wm}^{-1}\text{K}^{-1}$ or $\kappa_{\text{tan}}\,=\,116^{+7}_{-6}\,\text{Wm}^{-1}\text{K}^{-1}$. All $\kappa$ values that we derived by 2LRT in this work correspond to in-plane values ($\kappa_{\text{in-plane}}$) for our $c$-plane SM mainly made from GaN. Averaging over the two temperature gradients for each temperature cut orientation allowed us to obtain the stated, comparably low error intervals.  The sub- and superscript for our measured $\kappa_{\text{in-plane}}$ values denote the lower and upper bound of the asymmetric error interval. These asymmetries on the error intervals can be understood based on Eq.\,(\ref{eq:1}), relating $\kappa$ to the inverse of $\text{d}\,T\,/\,\text{d}\,\text{ln}\left( d / 1\,\mu m \right)$.

In general, we do not expect any intrinsic reasons for a thermal anisotropy in the $\textit{c}$-plane of our SMs. Thus, we assume that the lower value of $\kappa_{\text{rad}}$ compared to $\kappa_{\text{tan}}$ (difference of around 20$\%$) is linked to radial etch channels that can form in our SMs due to the electrochemical etching. Such channels can form at the perimeter of the central hole in the SM and progress from there on radially. Especially natural diamond is known to exhibit such etch channels.\cite{lu_observation_2001} However, recently Zhang \textit{et\,al.} also reported on the formation of etch channels in GaN due to electrochemical etching.\cite{zhang_study_2022} The observation of such thermal anisotropies based on temperature maps directly shows one of the key benefits of 2LRT compared to 1LRT measurements. Nevertheless, it remains an open research question to further analyze thermal transport by 2LRT in porous GaN or GaN material with aligned etch channels.

The maximal temperature rise ($T_{\text{rise}}$) interval that is used for the linear fitting of the data in Fig.\,\ref{fig:2LRT}(b) amounts to $\approx\,100\,\text{K}$. Remaining at such comparably low values of $T_{\text{rise}}$ for 2LRT measurements based on the Raman mode shift (a) guarantees a negligible impact of thermally induced stress. It was shown that such thermally induced stress can lead to significant deviations regarding the temperature determination, \cite{elhajhasan_optical_2023} if one either relies on Raman mode shifts (a) or Raman mode broadenings (b). In a first order approach, the Raman mode broadenings are stress-independent, contrary to the Raman mode shifts.\cite{callsen_phonon_2011} As a result, starting at $T_{\text{rise}}\,\approx\,150\,\text{K}$ in GaN one can observe a deviation of the temperatures extracted by either method (a) or (b). The local thermal expansion in the focus spot of the heating laser evokes a local compression of the sample due to thermal expansion, which most commonly leads to a shift of the Raman modes to higher energies.\cite{callsen_phonon_2011} This shift towards higher energies can in-turn be overcompensated by the shift of the Raman modes to lower energies due to local heating. As a result, we can observe artificially lowered $T_{\text{rise}}$ values in, e.g., GaN for $T_{\text{rise}}\,\gtrsim\,150\,\text{K}$, leading to an overestimation of $\kappa_{\text{in-plane}}$, cf. Eq.\,(\ref{eq:1}). In this work we circumvent such overestimations by imposing the aforementioned limit to $T_{\text{rise}}$.

\subsection{\label{subsec:Pabs}Determination of the absorbed power $P_{\text{abs}}$}

The absorbed laser power $P_{\text{abs}}$ is a very critical parameter for the determination of $\kappa_{\text{in-plane}}$ based on the 2LRT measurement from Fig.\,\ref{fig:2LRT}. At this point it still remains to be clarified how much of the entire absorbed laser power leads to the heating given by $P_{\text{heat}}$ that we observe for our SM. Commonly, the laser power $P_{\text{laser}}$ that impinges on the surface of our samples is split as follows:

\begin{equation}
\label{eq:2}
P_{\text{laser}}=P_{\text{abs}}+P_{\text{ref}}+P_{\text{trans}}+P_{\text{scat}}.
\end{equation}

A certain fraction of the laser power $P_{\text{laser}}$ is reflected ($P_{\text{ref}}$) or transmitted ($P_{\text{trans}}$), while pronounced light scattering with a power $P_{\text{scat}}$ can also occur. In general, $P_{\text{laser}}$ can simply be measured with a conventional laser power meter, which also allows measuring the transmission of all optical components in the laser beam path.

Subsequently, also $P_{\text{ref}}$ can be measured in our experimental setup for every value of $P_{\text{laser}}$ at $\lambda_{\text{heat}}$. For all $P_{\text{laser}}$ values used in this work we always observed a linear scaling with $P_{\text{ref}}$, which strongly suggests that this observation is also valid for $P_{\text{abs}}$. In addition, our measured value for $P_{\text{ref}}$ at $\lambda_{\text{heat}}$ agrees with reflectivity measurements based on a dedicated UV-VIS spectrophotometer and straightforward calculations of the reflectivity via the Fresnel equations for an angle distribution given by the NA of the microscope objective in use, cf. Sec.\,\ref{subsubsec:Raman}. As a result, we obtain a reflectivity of around $20\,\%$ at $\lambda_{\text{heat}}\,=\,325\,\text{nm}$.

The total thickness of our SM amounts to $830\,\text{nm}$, cf. Fig.\,\ref{fig:sample}. For determining $P_{\text{trans}}$ it is a good approximation to assume a pure GaN SM, which yields a transmission of $\approx\,0.001\,\%$ at $\lambda_{\text{heat}}\,=\,325\,\text{nm}$.\cite{kawashima_optical_2024} Thus, due to the high absorption of GaN at 325\,nm one can safely neglect $P_{\text{trans}}$ for the given thickness of the SM. Consequently, any possible interference effects in the SM \cite{elhajhasan_optical_2023} can be neglected, because the high absorption coefficient of GaN at 325\,nm ($\alpha\,\approx\,135000\,\,\text{cm}^{-1}$) translates to a light penetration depth of just 74\,nm.\cite{kawashima_optical_2024} Thus, any corrections related to $P_{\text{trans}}$ are negligible in the light of the error bars of our $\kappa$ values on the order of 5\,-\,15$\%$, cf. Sec.\,\ref{subsec:HRmaps}.

A similar statement is valid for $P_{\text{scat}}$. For the measurements of $P_{\text{ref}}$ we probe the specular reflection of the heating laser from the surface of our SM, which does not show any pronounced contribution of diffusely scattered light at deviating angles. We assign this observation to the monocrystallinity of our samples, which is accompanied by a low surface roughness (see Sec.\,\ref{subsec:roughness}) and a comparably low density of structural defects (e.g., dislocation density of a few 10$^{8}$\,cm$^{-2}$).\cite{Haller2017} Thus, we can safely neglect any contributions related to $P_{\text{scat}}$, which simplifies Eq.\,(\ref{eq:2}) to:

\begin{equation}
\label{eq:3}
P_{\text{laser}} \approx P_{\text{abs}} + P_{\text{ref}}\,=\,P_{\text{heat}} + P_{\text{rad}} + P_{\text{ref}}.
\end{equation}

\begin{figure*}[]
    \includegraphics[width=\linewidth]{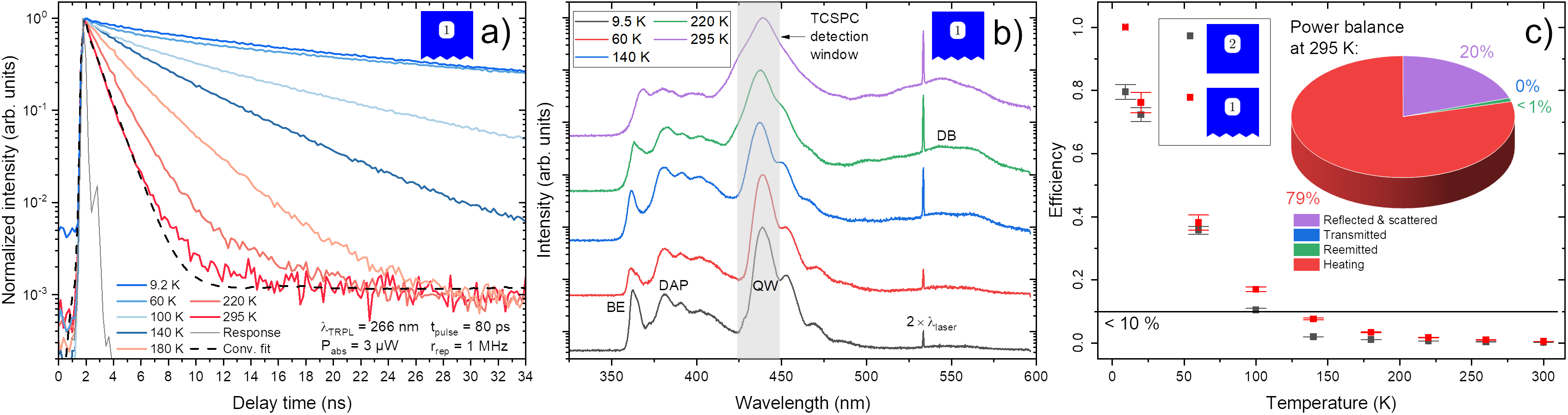}
    \caption{Photoluminescence characterization: (a) TRPL analysis of sample 1 showing transients related to the QW emission at various temperatures ($9.2\,\text{K}$\,-\,$300\,\text{K}$). The impact of non-radiative centers lowers the measured lifetimes, yielding, so-called, effective lifetimes $\tau_{\text{eff}}$. In addition, the temporal system response (gray solid line) and an example for the convoluted fit (black dashed line) are shown. (b) Corresponding PL spectra show the GaN bandedge (BE), the quantum well (QW), and the defect band (DB) emission of sample 1 for selected temperatures. In addition, the donor-acceptor-pair (DAP) emission is observed at $9.5\,\text{K}$. The gray-shaded area illustrates the detection window that formed the basis for the recorded transients. (c) Estimation of the PL efficiency of samples 1 and 2 for temperatures up to $295\,\text{K}$ under excitation with a pulsed laser emitting at $\lambda_{\text{TRPL}}\,=\,266\,\text{nm}$. Please note that this wavelength $\lambda_{\text{TRPL}}$ provides an excitation of GaN above its bandgap similar to the heating laser used for 2LRT measurements. The pie chart summarizes the laser-induced power balance.}
    \label{fig:tart}
\end{figure*}

Consequently, we have to determine the fraction of the laser power that is re-emitted by our samples, which we denote by $P_{\text{rad}}$. Due to the comparably lower values of $T_{\text{rise}}$ and $T_{\text{amb}}$ in this work and the rather large gap of $\approx\,100\,\text{nm}$ between our SM and the underlying nitride material, we can neglect thermal radiation. However, our SMs are photonic membranes, which means that their internal structure based on a stack of QWs is made to output light. The main contributions of the related PL signal arises from a temperature-dependent balance of band-to-band transitions and excitonic transitions in the QWs, which are accompanied by a larger variety of other possible origins of PL signal in III-nitrides detailed in the following. Therefore, to estimate the overall re-emitted power of our sample $P_{\text{rad}}$, we perform time-resolved PL spectroscopy as summarized in Fig\,\ref{fig:tart}.

The most pronounced source of PL signal in our samples are the embedded In$_{0.15}$Ga$_{0.85}$N QWs. Therefore, Fig.\,\ref{fig:tart}(a) shows the decay dynamics of the QW emission for temperatures ranging between 9.2\,K and 295\,K. We observe a strong decrease of the measured lifetimes with increasing temperature due to the increasing contribution of non-radiative defects. We employ a convoluted fitting routing for all PL transients, which is exemplified for $T_{\text{amb}}\,=$\,295\,K by the black dashed line, cf. Fig\,\ref{fig:tart}(a). For this fitting procedure we convolute the measured temporal response of our detection system (see the gray solid line) with a single exponential decay. The resulting dataset is fitted to the PL decay curves by altering the essential parameters of the exponential function such as the effective decay time $\tau_{\text{eff}}$, which represents a standard procedure for such an analysis.\cite{cross_analysis_1984} The spectral window that was covered during the recording of the PL transients related to the QW emission is illustrated by the gray-shaded area in Fig.\,\ref{fig:tart}(b). Here, we display the entire PL spectrum of sample 1 from the UV to the visible spectral range for some selected temperatures. At a temperature of 9.2\,K the PL spectrum starts (from left to right) with the bandedge (BE) emission, continues with the donor-acceptor-pair emission (DAP) and the main QW emission with its longitudinal-optical (LO) phonon replicas, before a broad defect luminescence band (DB) appears at around 500\,-\,600\,nm. All these different sources of PL are well-documented in literature for $p$-type GaN \cite{callsen_optical_2012} that forms the top layer of our entire layer stack, cf. Fig.\,\ref{fig:sample}(a). It shall be noted that the DAP emission gradually transitions to a band-acceptor emission (e$^{-}$A) starting at at temperatures $\approx\,60\,\text{K}$ due to the increasing ionization of all involved donor states. \cite{callsen_optical_2012} For the following analysis of Eq.\,(\ref{eq:3}) we can focus on the main source of PL that has a dominating contribution to $P_{\text{rad}}$ - the QW emission band.

Following a procedure described by Langer \textit{et\,al.}, \cite{langer_room_2013} one can estimate the internal quantum efficiency (IQE) of a QW for all temperatures by measuring the temperature dependence of $\tau_{\text{eff}}(T)$ and the initial QW PL intensity $I_{0}(T)$ at the beginning of the temporal decay depicted in Fig.\,\ref{fig:tart}(a). Commonly, resonant or quasi-resonant excitation \cite{weatherley_imaging_2021} of the QW emission is required for such a determination of the IQE. However, such sample excitation below the absorption edge of GaN is not compatible with our case of an effective laser-induced heating of a photonic SM with $\lambda_{\text{heat}}\,=\,325\,\text{nm}$. Thus, we performed our TRPL analysis with a sufficiently short laser wavelength of $\lambda_{\text{TRPL}}\,=\,266\,\text{nm}$ yielding a light penetration depth of $\approx$\,45\,nm in GaN.\cite{kawashima_optical_2024} As a result, carrier transfer from the 150-nm-thick $p$-type GaN cap layer of our sample to the QWs becomes essential, which means that the TRPL analysis is influenced by non-radiative loss mechanisms in the matrix material (GaN) \textit{and} the QW. This is especially relevant as the top $p$-type GaN layer is thicker than the light penetration depth of our heating laser (74\,nm at $\lambda_{\text{heat}}\,=\,325\,\text{nm}$) and the laser used for the TRPL measurements ($\approx$\,45\,nm at $\lambda_{\text{TRPL}}\,=\,266\,\text{nm}$).  Thus, our measurements do not result in an estimate of the sample IQE, but an "efficiency" of the dominating emission for the given excitation conditions required for sample heating. The corresponding results for samples 1 and 2 are plotted in Fig.\,\ref{fig:tart}(c). It shall be noted that these results represent an upper boundary estimation of $P_{\text{rad}}$, because $\tau_{\text{eff}}(T)$ and $I_{0}(T)$ are only linked to the dominating QW emission. When considering some of the other aforementioned sources of PL emission (BE, DAP, e$^{-}$A) in our samples, one obtains even lower efficiency values because the QW emission proves to be more robust with rising temperature. Here, only the DB represents an exception, because it is clearly activated with rising temperature as shown in Fig.\,\ref{fig:tart}(b). Nevertheless, its overall intensity remains small compared to the emission of the QWs. In general, all these additional PL contributions only yield minor corrections to our estimation of the sample efficiency and therefore $P_{\text{rad}}$, because the QW emission dominates the PL spectra at all relevant temperatures up to 295\,K.

Thus, any relevant contributions of $P_{\text{rad}}$ to our determination of $\kappa_{\text{in-plane}}$ only occur for $T_{\text{amb}}\,\leq\,100\,\text{K}$ as shown in Fig.\,\ref{fig:2LRT}(c). For all more elevated temperatures the corrections given by $P_{\text{rad}}$ compare-well to the error of $\kappa_{\text{in-plane}}$, which is commonly on the order of 5\,-15\,$\%$ for 2LRT measurements. \cite{elhajhasan_optical_2023}. Nevertheless, in the present work that focuses on $T_{\text{amb}}\,=\,295\,\text{K}$, we still consider all corrections to $P_{\text{abs}}$ by $P_{\text{rad}}$ (see Eq.\,(\ref{eq:3})). However, because $P_{\text{rad}}$ only amounts to $\approx\,0.5\%$ of $P_{\text{laser}}$, which is also small compared to $P_{\text{heat}}$, one could safely simplify the right-hand side of Eq.\,(\ref{eq:3}). As a result, we obtain $P_{\text{abs}}\,\approx\,P_{\text{heat}}$ for our samples as long as $T_{\text{amb}}\,>\,100\,\text{K}$ holds. Thus, the consideration of $P_{\text{rad}}$ will only play a dominant role for thermal transport studies on GaN membranes by 2LRT at cryogenic temperatures. In addition, the consideration of $P_{\text{rad}}$ can even be needed at higher temperatures depending on the sample quality that can further be improved, e.g., by transitioning to native GaN substrates and/or the application of, so-called, indium-containing underlayers that can boost the IQE of InGaN QWs.\cite{Haller2017} Especially for thermometry on photonic membranes it always remains to carefully check the relevance of $P_{\text{rad}}$ for achieving a meaningful determination of $P_{\text{heat}}$.

Finally, the entire power balance of $P_{\text{ref}}$, $P_{\text{scat}}$, $P_{\text{trans}}$, $P_{\text{rad}}$, and $P_{\text{heat}}$ at $T_{\text{amb}}\,=295\,\text{K}$ can be plotted for our samples as shown in the inset of Fig.\,\ref{fig:tart}(c). Almost 80$\%$ of the incident laser power leads to a heating of our samples, which is key to the determination of $\kappa_{\text{in-plane}}$. Any 2LRT analysis of even brighter samples with higher $P_{\text{rad}}$ at $T_{\text{amb}}\,=\,$295\,K would further be challenged by the PL background in the Raman spectra as exemplified in Fig.\,\ref{fig:sample}(d). Under the high excitation conditions at the given high values of $P_{\text{abs}}$ that are required for 2LRT measurements (see Fig.\,\ref{fig:2LRT}) one can observe a broadening of the QW emission,\cite{Nippert2016a} which leads to a PL tail that, e.g., can appear as a background in the Raman spectra. In addition, the aforementioned, thermally activated defect band appears around $\lambda_{\text{probe}}\,=\,532\,\text{nm}$ as shown in Fig.\,\ref{fig:sample}(b), troubling the acquisition of Raman spectra. Therefore, it would be beneficial to shift $\lambda_{\text{probe}}$ above 600\,nm. However, the commercially available microscope objectives limit such possibilities. Commonly available, so-called, two- and three-wavelengths microscope objectives are only optimized for certain harmonics of Nd:YAG lasers at, e.g., 266\,nm, 355\,nm, and 532\,nm. Thus, for future work it would be promising to tune $\lambda_{\text{probe}}$ to wavelength ranges that correspond to PL minima indicated by the PL spectra of Fig.\,\ref{fig:tart}(b). In addition, it also still remains a task for future research to vary $P_{\text{abs}}$ during the TRPL measurements to check whether an increase of the PL efficiency can be achieved by saturating non-radiative defects in GaN that hinder the carrier transfer to the QWs. Nevertheless, the interpretation of such measurements will prove challenging due to the fundamentally different excitation schemes of 2LRT and TRPL measurements, which either rely on a pulsed or a cw laser. Furthermore, a partial screening of the built-in electric fields will also occur upon the increase of $P_{\text{abs}}$ in the In$_{\text{0.15}}$Ga$_{\text{0.85}}$N/GaN heterostructure of our SMs, \cite{schlichting_suppression_2018} which will further complicate the analysis. In a best case scenario, the determination of the sample PL efficiency would be performed with the same laser and the same laser powers that are used during the 2LRT measurements. Thus, corresponding efficiency measurements in an integration sphere over wide temperature ranges (from cryogenic temperatures up to the highest temperatures achieved during 2LRT) appears as a promising task for future work.

\subsection{\label{subsec:roughness}Thermal imaging and the impact of surface roughness}

\begin{figure*}[]
    \includegraphics[width=\linewidth]{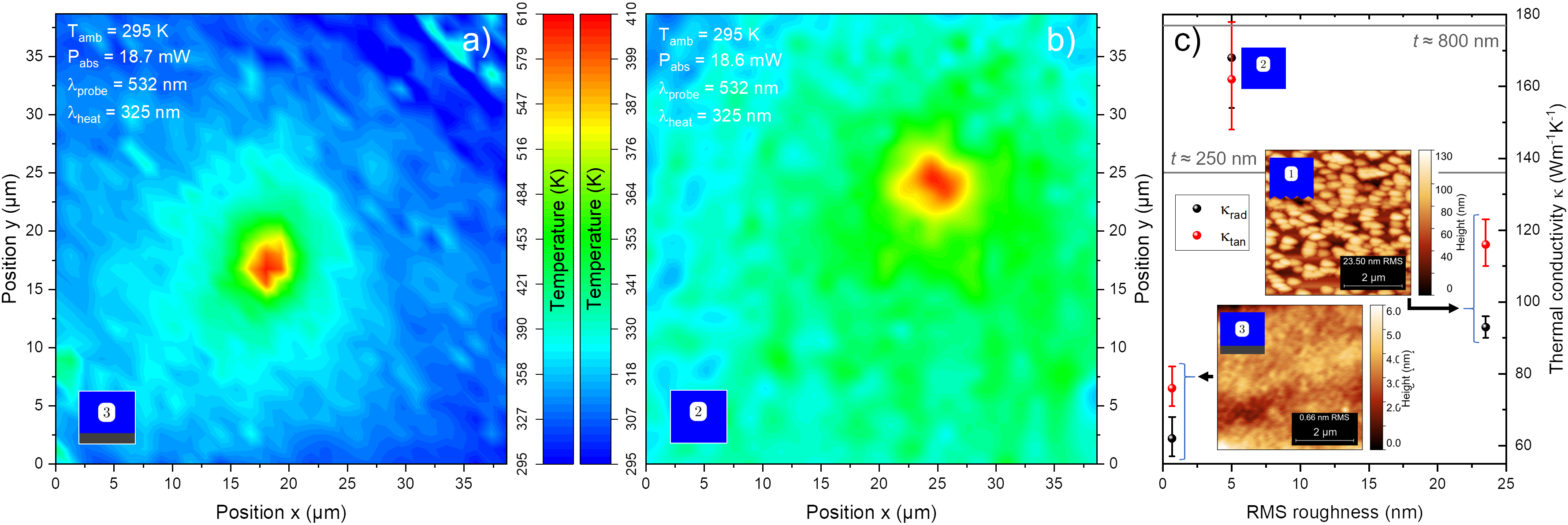}
    \caption{Impact of backside roughness on $\kappa$: (a) Temperature mapscan of sample 3 obtained by 2LRT measurements (step resolution $\approx\,1.3\,\mu\text{m}$). (b) Corresponding temperature mapscan of sample 2 obtained with similar experimental parameters. The different $\kappa$ values of samples 2 and 3 become directly visible by comparing the maximal temperature rise and the overall temperature distribution (very localized vs. more extended heating). (c) Summary of all radial ($\kappa_{\text{rad}}$) and tangential ($\kappa_{\text{tan}}$) values for the thermal conductivity obtained for samples 1\,-\,3. Here, the stated route mean square (RMS) values for the roughness of the backside of the photonic membranes are supported by two AFM images linked to samples 1 and 3. The horizontal gray lines show our theoretical $\kappa$ values for two exemplary membrane thicknesses.}
    \label{fig:roughness}
\end{figure*}

Our 2LRT imaging does not only provide access to the thermal conductivity $\kappa$ of our SMs for all in-plane directions. Major differences of $\kappa_{\text{in-plane}}$ can even be made directly visible as highlighted in Fig.\,\ref{fig:roughness}. Sufficiently large $\kappa_{\text{in-plane}}$ differences can be generated by tuning the backside surface roughness of our samples, while the frontside surface roughness remains constant. So far sample 1 was analyzed in Fig.\,\ref{fig:2LRT}(a), which showed the roughest backside surface morphology due to a processing bias voltage of 10\,V, cf. Sec.\,\ref{subsec:Sample}. A corresponding AFM image is shown in the inset of Fig.\,\ref{fig:roughness}(c), yielding a root mean square (RMS) roughness of 23.5\,nm obtained from a 5\,$\times$\,5\,$\mu \text{m}^{2}$ scan. Decreasing this processing voltage to 6\,V reduces this RMS roughness to 5.0\,nm for sample 2 (same scan size). The corresponding AFM image of the backside of sample 2 can be found in the Supplemental Material in S-Sec.\,I. Any further decrease of the backside roughness of our samples could only be achieved by implementing a polarization field in the sacrificial layer via an additional AlN/AlInN interlayer (1\,nm of AlN (nid) and 20\,nm of Al$_{0.82}$In$_{0.18}$N). \cite{ciers_smooth_2021} It shall be noted that such variation in the bottom interface does not impact the core region of the SM where the QW is positioned, meaning that our approximation of $P_{\text{rad}}$ from Sec.\,\ref{subsec:Pabs} remains valid for all of our samples. As a result, sample 3 exhibits an RMS backside roughness of just 0.66\,nm as shown in the inset of Fig.\,\ref{fig:roughness}(c), which is almost identical to the frontside RMS roughness of all epilayers (frontside RMS of sample 1 and 2: 0.67\,nm, sample 3: 0.76\,nm) in use.\cite{ciers_smooth_2021}

However, the addition of the AlN/AlInN interlayer to sample 3 comes at the cost of strongly reduced $\kappa_{\text{in-plane}}$ values as summarized in Fig.\,\ref{fig:roughness}(c) for samples 1\,-\,3 for all radial ($\kappa_{\text{rad}}$) and tangential ($\kappa_{\text{tan}}$) in-plane directions with respect to the hole in the SMs, cf. Fig.\,\ref{fig:2LRT}(a). The values of $\kappa_{\text{rad}}$ and $\kappa_{\text{tan}}$ for sample 3 are more than halved compared to sample 2 that exhibits a more than five times higher RMS backside roughness. Thus, the backside termination of sample 3 with AlN and Al$_{0.82}$In$_{0.18}$N does not prove beneficial from a thermal point of view. The additional interfaces apparently boost the phonon scattering and consequently reduce $\kappa$. In addition, the AlN and Al$_{0.82}$In$_{0.18}$N layers provide large acoustic contrast compared to GaN, yielding a higher thermal resistance due to the mismatch of the phonon dispersion relations of all materials that contribute to the SM of sample 3.

Interestingly, the pronounced difference between all $\kappa$ values related to samples 3 and 2 becomes visible when comparing Figs.\,\ref{fig:roughness}(a) and \ref{fig:roughness}(b). For almost identical values of $P_{\text{abs}}$ one reaches maximal temperatures $T_{\text{max}}$ in the central heat spot that differ by around 200\,K. Further direct visual investigation of the two thermal images in Fig.\,\ref{fig:roughness} yields that the heat spot for sample 3 is much more localized compared to sample 2, which is a direct sign of lower $\kappa_{\text{in-plane}}$ values. It shall be noted that the different temperature ranges in Figs.\,\ref{fig:roughness}(a) and \ref{fig:roughness}(b) are required for the illustration of the thermal images due to the strongly deviating $T_{\text{max}}$ values. Overall, 2LRT measurements become increasingly challenging for high values of $\kappa_{\text{in-plane}}$, because the heated area starts to extend over several tens of micrometers. In general, it is not required to encompass the entire temperature gradient between $T_{\text{max}}$ and $T_{\text{amb}}$ in a 2LRT image, because the determination of $\kappa_{\text{in-plane}}$ is only based on temperature differences given by $T_{\text{rise}}$. However, higher values of $\kappa_{\text{in-plane}}$ go hand-in-hand with lower values of $T_{\text{rise}}$, which are increasingly difficult to resolve by Raman thermometry.\cite{elhajhasan_optical_2023} In addition, as mentioned in Sec.\,\ref{subsec:HRmaps}, we aim to limit the $T_{\text{rise}}$ values to mitigate the impact of thermally induced stress for our 2LRT analyses.

Already based on the data from Fig.\,\ref{fig:2LRT} we observed that $\kappa_{\text{rad}}$ was by around 20$\%$ smaller compared to $\kappa_{\text{tan}}$ for sample 1. This anisotropy, which is most likely linked to radial etch channels in our SMs close to the central hole as mentioned in Sec.\,\ref{subsec:HRmaps}, reappears for sample 3 but is absent for sample 2, cf. Fig.\,\ref{fig:roughness}(c). Interestingly, sample 2 exhibits the highest $\kappa_{\text{in-plane}}$ values, which underlines the detrimental impact of such etch channels for thermal transport. 

\subsection{\label{subsec:modeling}\textit{Ab initio} modeling of the thermal conductivity $\kappa$}

We compute $\kappa_{\text{in-plane}}$ via an \textit{ab initio} solution of the linearized phonon Boltzmann transport equation (BTE) as implemented in the \texttt{ELPHBOLT} package.\cite{protik_elphbolt_2022} The full density functional theory based workflow has previously been detailed in Ref. \onlinecite{elhajhasan_optical_2023}. From the phonon mode (identified by the band $s$ and the wave vector $\mathbf{q}$) resolved solution of the BTE we can, e.g., calculate the phonon mean free path distribution and the thermal conductivity tensor of wurtzite GaN. We consider here both 3- and 4-phonon scattering. The latter proves to be relevant for wurtzite GaN due to a pronounced gap separating the high-lying optic sector from the rest in the phonon band structure and a consequent dip in the 3-phonon scattering phase space.\cite{elhajhasan_optical_2023} In addition, the phonon-isotope scattering is considered within the Tamura model, \cite{tamura_isotope_1983} while the mode-resolved phonon-boundary scattering rates are computed using the following empirical expression:

\begin{equation}
\label{eq:4}
\dfrac{1}{\tau_{s\mathbf{q}}^{\text{phonon-boundary}}} = \frac{2 v_{s\mathbf{q}}^{\perp}}{t},
\end{equation}

with the mode-resolved phonon group-velocity components perpendicular to the plane of the SM ($v_{s\mathbf{q}}^{\perp}$) and the membrane thickness ($t$). Equation\,(\ref{eq:4}) considers fully diffusive scattering of phonons at the top and bottom of our photonic SMs. As a result, we obtain $\kappa_{\text{in-plane}}\,=\,177\,$Wm$^{-1}$K$^{-1}$ for an around 800-nm-thick pure GaN membrane, which compares well to our results for sample 2, yielding $\kappa_{\text{rad}}\,=\,168^{+15}_{-14}\,$Wm$^{-1}$K$^{-1}$ and $\kappa_{\text{tan}}\,=\,162^{+16}_{-14}\,$Wm$^{-1}$K$^{-1}$. The corresponding theoretical cumulative $\kappa_{\text{in-plane}}$ and its dependence on $l_{\text{mfp}}$ is plotted in the Supplemental Material in S-Sec.\,II for two exemplary SM thicknesses $t$. We also discuss the applied averaging process (isotropic vs. pure in-plane averages) for the determination of our theoretical $\kappa$ values in S-Sec.\,II. Figure\,\ref{fig:roughness}(c) comprises two horizontal gray lines that summarize the results of our modeling of $\kappa_{\text{in-plane}}$ for SM thicknesses ($t$) around 250\,nm and 800\,nm.

The impact of the different layers in the layer stack of samples 1 and 2 (e.g., GaN layers with different doping, QWs) seems almost negligible compared to the addition of the etch stop layers made from AlN and Al$_{0.82}$In$_{0.18}$N for sample 3. A brief inspection of the layer stack related to samples 1 and 2 as shown in Fig.\,\ref{fig:sample}(a) yields that almost the entire photonic SM is made from GaN in agreement with our numerical simulations. The $n$- and $p$-type doping of GaN up to $3\,\times\,10^{18}\,\text{cm}^{-3}$ ($n_{\text{Si}}$) and $1\,-\,5\,\times\,10^{19}\,\text{cm}^{-3}$ ($n_{\text{Mg}}$) does not significantly reduce $\kappa_{\text{in-plane}}$ \cite{elhajhasan_optical_2023} due to the low atomic fraction. The main acoustic contrast in the material stack of samples 1 and 2 arises from the multi-QW stack. However, this stack comprises only three 2.8-nm-thick In$_{0.15}$Ga$_{0.85}$N QWs. As a result, only $1\%$ of our photonic SM consists of material that is not GaN and even in this case the atomic fraction of foreign atoms is small with just $15\%$. Clearly, this justifies our theoretical approach and further explains the good agreement between the theoretical and experimental results for sample 2. By a similar line of reasoning we can understand why the addition of the AlN and Al$_{0.82}$In$_{0.18}$N layers has such a strong impact on $\kappa_{\text{in-plane}}$ for sample 3. Both layers are also thin with 1\,nm and 20\,nm, however, the atomic composition deviates completely, leading to a large acoustic contrast for simple mass reasons. Nevertheless, it represents an interesting task for future work to calculate the thermal impact of such thin additional layers that cannot only act as an etch stop, but could also be used for surface passivation purposes.\cite{chen_gan_2024}

\subsection{\label{subsec:mfp}The role of the phonon mean free paths}

Despite our detailed determination of $P_{\text{abs}}$ in Sec.\,\ref{subsec:Pabs}, it remains questionable over which volume heating occurs. Even though the laser spot size can be precisely measured (see Sec.\,\ref{subsubsec:2LRT}), the extent of the heated volume is unknown due to the distribution of $l_{\text{mfp}}$, which is characteristic for each sample. For instance, phonons with $l_{\text{mfp}}$ values even beyond, e.g., 1\,$\mu$m can significantly contribute to $\kappa$ in bulk GaN. \cite{elhajhasan_optical_2023} Thus, for our GaN SMs it is not sufficient to perform 1LRT measurements as soon as the laser spot diameter approaches $l_{\text{mfp}}$ values of phonons that exhibit a significant contribution to $\kappa$. It shall be noted that this statement remains also true when, e.g., relevant $l_{\text{mfp}}$ values only reach 10$\%$ of the laser spot diameter, because still phonon escape from the temperature probe volume can occur during 1LRT measurements. As a result, 1LRT measurements for which one laser acts as the heat source and the temperature probe, can yield too large $\kappa$ values for high quality samples.\cite{elhajhasan_optical_2023} A workaround to this problem would be an increase of the laser spot diameter during 1LRT measurements, however, this contradicts the aim of a truly local thermal characterization.

Therefore, 2LRT measurements can be beneficial by enabling the determination of temperature gradients with a distance of several micrometers to the central heat spot, cf. Fig.\,\ref{fig:2LRT}(b). In such regions our numerical modeling of the experimental data, based on \texttt{COMSOL} to solve the three-dimensional heat transport equation is more adequate. A next step would be the comparison of our experimental results to the output of a real-space solver for \textit{ab initio} phonon transport like OpenBTE.\cite{romano_openbte_2021} However, solving the steady-state phonon  BTE based on input from \textit{first-principles} calculations (e.g., based on \texttt{ELPHBOLT} \cite{protik_elphbolt_2022}) represents a numerical challenge for large scale structures of several tens of micrometers in size, comprising vertical boundaries (e.g., the boundary of the central hole in our membrane) and horizontal interfaces (e.g, the layer stack in the photonic SM). Thus, it remains as task for future work to first reduce the level of complexity by receding towards truly pure GaN SMs, before adding individual interfaces and boundaries in a step-by-step approach. Another promising approach that we plan to explore is the extraction of the $\kappa$ values via a differentiable BTE solver.\cite{romano_inverse_2022} In such a scenario, the thermal conductivity is derived from comparing simulated temperature maps, which build, e.g., on the full distribution of phonon $l_{\text{mfp}}$ values, with the experimental temperature maps derived from 2LRT. Nevertheless, already based on the present work it was shown that 2LRT is a promising tool at hand for such future studies, which proved to be sufficiently sensitive to probe, e.g., interface roughnesses and the addition of thin interlayers made from AlN and Al$_{0.82}$In$_{0.18}$N. As a next step it remains to proceed towards thermal imaging at $T_{amb}\,\leq\,295\,\text{K}$ for 2LRT to explore, e.g., different phonon transport regimes \cite{cepellotti_phonon_2015} and the overall rise of $l_{\text{mfp}}$ by imaging non-Fourier temperature distributions.\cite{beardo_hydrodynamic_2022}

\section{\label{sec:Conclusions}Conclusions}

In this work we demonstrated the non-invasive thermal characterization of photonic SMs that can, e.g., be applied for VCSEL structures. We showed how to cope with highly emissive samples during our all-optical thermal characterization by considering the fraction of light that is reemitted during the laser-induced heating. Therefore, we performed a TRPL analysis to derive a vital threshold for our thermometry of $T_{\text{amb}}\,\approx\,100\,\text{K}$. Below this threshold temperature, any neglect of the QW-induced light emission in our samples leads to an erroneous determination of $\kappa_{\text{in-plane}}$. Here, our work allowed us to point out the relevance of a detailed optical pre-characterization that is needed for an all-optical thermal characterization. Furthermore, we disentangled the power balance for our optical thermometry by considering the power fractions of the reflected, scattered, transmitted, and reemitted light. As a result, we obtained an estimate of the fraction of the laser power that is converted into heat, which is a general challenge for optical thermometry - especially if photonic structures are involved that are intended to emit light.

One highlight of our work is the thermal imaging by 2LRT over several tens of micrometers, which allows us to overcome the limits of conventional Raman thermometry based on just one laser acting as the heating and probe laser simultaneously. By our thermal imaging we found a thermal anisotropy in the plane of our photonic SMs, which is linked to extrinsic reasons such as etch channels in our SMs. Clearly, this shows how well-suited 2LRT measurements are for studying thermally anisotropic materials like, e.g., Ga$_{2}$O$_{3}$.\cite{jiang_three-dimensional_2018} By extended \texttt{COMSOL} simulations based on a three-dimensional solution of the heat transport equation, we highlighted how to extract $\kappa_{\text{in-plane}}$ values from 2LRT measurements beyond the limits of commonly applied analytical solutions of the same equation that, e.g., do not consider lateral sample boundaries.

In a best case scenario our thermal imaging renders differences in $\kappa_{\text{in-plane}}$ among our samples directly visually accessible to the experimentalist. Such differences were linked to variations of the SM backside roughness and additions to the corresponding layer stack. It was shown that despite the presence of three, just a few-nm-thick QWs in our samples, one can still approximate them as being made from pure GaN material. In contrast, the addition of interlayers made from AlN and Al$_{0.82}$In$_{0.18}$N reduces $\kappa_{\text{in-plane}}$ by more than a factor of two due to their large acoustic contrast compared to GaN. Finally, for an around 800-nm-thick pure GaN membrane we simulated $\kappa_{\text{in-plane}}\,=\,177\,$Wm$^{-1}$K$^{-1}$ ($c$-plane), which compares well to our experimental 2LRT results for our best sample with a comparable thickness, yielding an averaged in-plane $\kappa_{\text{in-plane}}$ of $165^{+16}_{-14}\,$Wm$^{-1}$K$^{-1}$.

The thermal analysis detailed in this work shall provide a pathway towards thermal imaging over wide temperature ranges, which is key to most direct experimental analyses of different phonon transport regimes or even the probing of phonon-polariton transport across interfaces.

\begin{acknowledgments}

N.H.P. acknowledges funding from the "Deutsche Forschungsgemeinschaft" (DFG, German Research Foundation) for an "Emmy Noether" research grant (Grant No. 534386252). G.R. acknowledges funding from the MIT-IBM Watson AI Laboratory (Challenge No. 2415). G.R. and G.C. acknowledge funding from the MIT Global Seed Funds. M.E. and G.C. acknowledge funding from the Central Research Development Fund (CRDF) of the "Universität Bremen" for the project "Joint optical and thermal designs for next generation nanophotonics". The research of W.S., M.E., K.D., G.W., and G.C. was further funded by the major research instrumentation program of the DFG (Grant No. 511416444). G.C. further acknowledges the MAPEX-CF Grant for Correlated Workflows (Grant No. 40401080). J.C. and \AA.H. acknowledge the financial support from the European Research Council (ERC) under the European Union’s Horizon 2020 Research and Innovation Program (Grant Agreement No. 865622), and the Swedish Research Council (Grant No. 2018-00295). The samples were processed at Myfab Chalmers University of Technology.

\end{acknowledgments}
%
%
%








\section*{Data Availability Statement}

The data that support the findings of this study are available from the corresponding author upon reasonable request.

\bibliography{AA)GaN_membrane1}

\end{document}


\title[]{SUPPLEMENTAL MATERIAL: \\Thermal analysis of GaN-based photonic membranes for optoelectronics}

\author{Wilken Seemann}
\affiliation{Institut für Festk\"orperphysik, Universit\"at Bremen, Otto-Hahn-Allee 1, 28359 Bremen, Germany}
\altaffiliation[Wilken Seemann also at:]{\,MAPEX Center for Materials and Processes, Universität Bremen, \\ Bibliotheksstraße 1, 28359 Bremen, Germany}

\author{Mahmoud Elhajhasan}
\author{Julian Themann}
\author{Katharina\,\,Dudde}
\author{Guillaume W\"ursch}
\author{Jana Lierath}
\author{Gordon Callsen}
\email{gcallsen@uni-bremen.de}

\affiliation{Institut für Festk\"orperphysik, Universit\"at Bremen, Otto-Hahn-Allee 1, 28359 Bremen, Germany}

\author{Joachim Ciers}
\author{\AA sa Haglund}

\affiliation{Department of Microtechnology and Nanoscience, Chalmers University of Technology, 41296 Gothenburg, Sweden}

\author{Nakib H. Protik}

\affiliation{Department of Physics and CSMB, Humboldt-Universit\"at zu Berlin, 12489 Berlin, Germany}

\author{Giuseppe Romano}

\affiliation{MIT-IBM Watson AI Lab, IBM Research, Cambridge, MA, USA}

\author{Rapha\"el Butt\'{e}}
\author{Jean-François Carlin}
\author{Nicolas Grandjean}

\affiliation{Institute of Physics, \'{E}cole Polytechnique F\'{e}d\'{e}rale de Lausanne (EPFL), CH-1015 Lausanne, Switzerland}


\date{\today}

\maketitle

\section{\label{S-Sec:Calibration}Atomic force microscopy on sample 2}

\begin{figure*}[h]
    \includegraphics[width=0.5\linewidth]{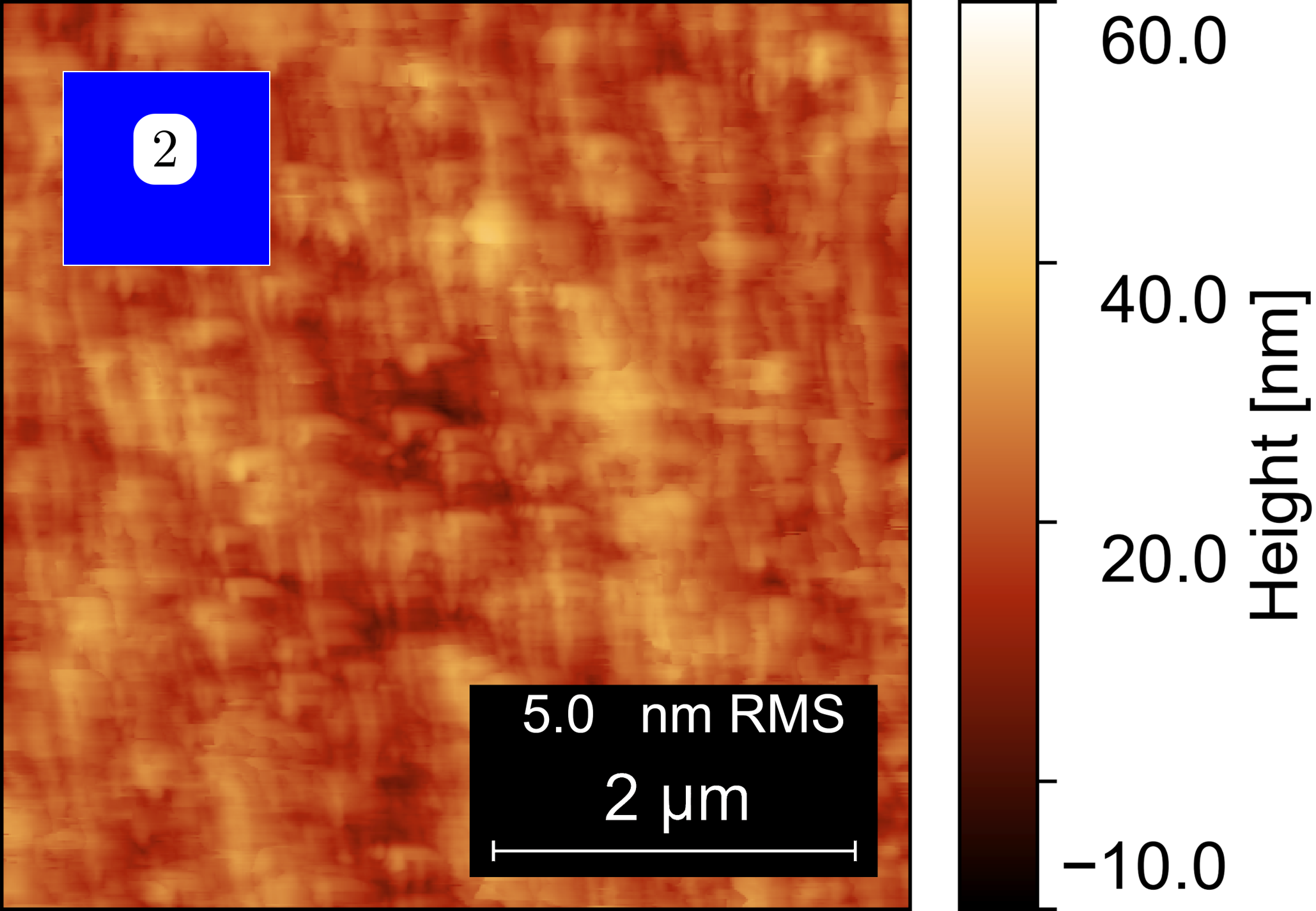}
    \caption{\textit{AFM image taken on the backside of sample 2 leading to an RMS surface roughness of 5.0\,nm. The blue symbol in the inset that indicates sample 2 is introduced in Fig.\,1(a) \newline of the main text.}}
   \label{S-Fig:AFM}
\end{figure*}

 S-Figure\,\ref{S-Fig:AFM} shows an image of the backside of sample 2 that was obtained by atomic force microscopy (AFM). From this image that is 5\,$\times$\,5\,$\mu \text{m}^{2}$ in size, one obtains a root mean square (RMS) surface roughness of 5.0\,nm. This RMS roughness falls between the corresponding roughness of sample\,3 (0.66\,nm) and that of sample 1 (23.5\,nm). The AFM image from S-Fig.\,\ref{S-Fig:AFM} complements the AFM image of Fig.\,4(c) of the main text.

%
%
%

\section{\label{S-Sec:Etchfronts}Modeling of the cumulative thermal conductivity}

\begin{figure*}[]
    \includegraphics[width=0.5\linewidth]{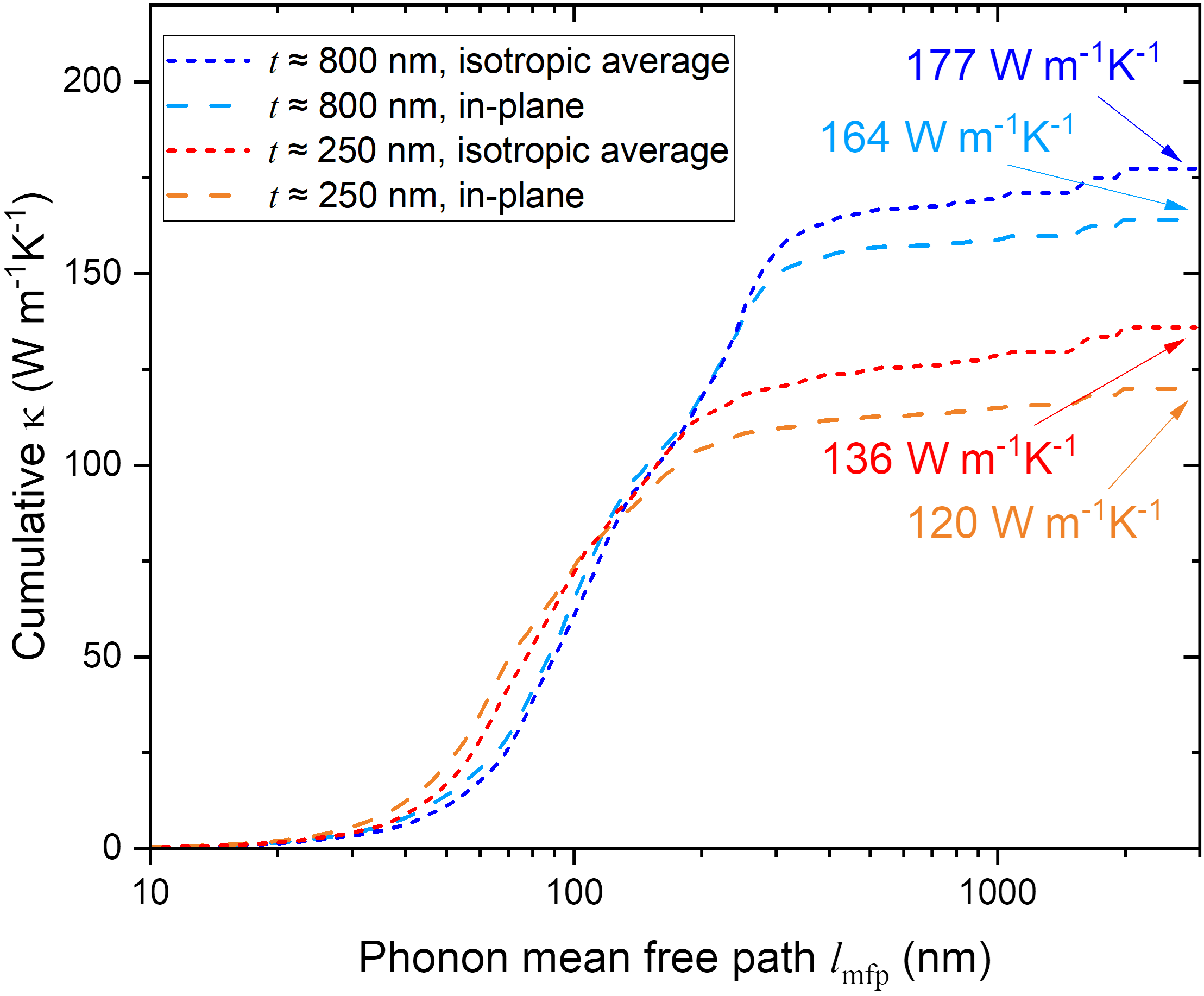}
    \caption{\textit{Accumulation of the phonon thermal conductivity $\kappa$ with respect to the phonon mean free path ($l_{\text{mfp}}$) for an around $250$-nm- (red color group) and $800$-nm-thick (blue color group) semiconductor membrane (SM) made from pure wurtzite, $c$-plane GaN. For each SM thickness we also compare the isotropic average of $\kappa$ with its in-plane value $\kappa_{\text{in-plane}}$.}}
    \label{fig:kappa_accum_mfp}
\end{figure*}

\begin{figure*}[h]
    \includegraphics[width=0.75\linewidth]{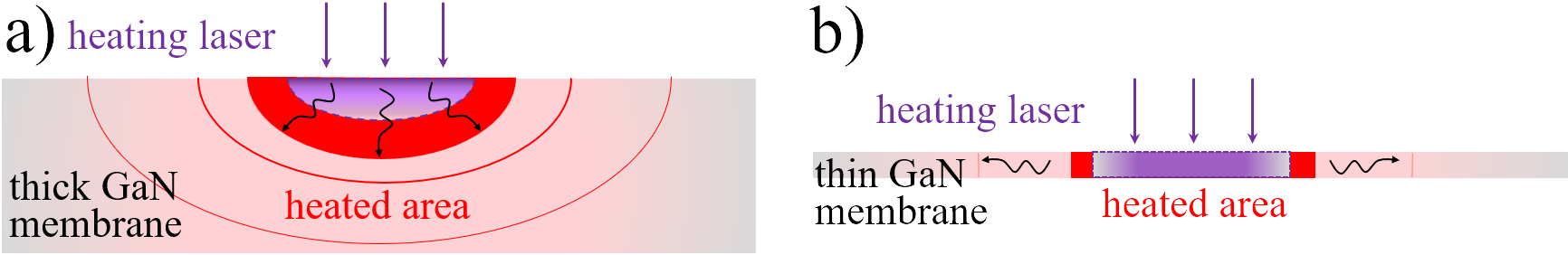}
    \caption{\textit{Sketch illustrating the ratio between in- and cross-plane thermal transport in a thick (a) and a thin (b) GaN membrane as a result of heating provided by a heating laser. This ratio depends on the fraction of the membrane thickness that is heated by the heating laser.}}
    \label{fig:surfacevolumeheating}
\end{figure*}

In S-Fig.\,\ref{fig:kappa_accum_mfp} we illustrate the thermal conductivity $\kappa$ accumulation with respect to the phonon mean free path (mfp) $l_{\text{mfp}}$ for two exemplary semiconductor membranes (SMs) made from pure wurtzite $c$-plane GaN. From S-Fig.\,\ref{fig:kappa_accum_mfp}, we obtain $\kappa_{\text{in-plane}}\,=\,177\,$Wm$^{-1}$K$^{-1}$ for an $\approx 800$-nm-thick SM made from wurtzite $c$-plane GaN. In addition to $\kappa_{\text{in-plane}}$, we also plot the isotropic average $\kappa_{\text{isotropic}}$ $= (2\kappa_{\text{in-plane}} + \kappa_{\text{cross-plane}})/3$ in S-Fig.\,\ref{fig:kappa_accum_mfp}, which yields smaller overall values. This is caused by the fact that the bulk $\kappa$ values of wurtzite GaN exhibit a small anisotropy (in-plane vs. cross-plane), compared to the anisotropy that is induced by the phonon-boundary scattering occurring in our $c$-plane SMs. Interestingly, the resulting value of $\kappa_{\text{isotropic}}\,=\,164\,$Wm$^{-1}$K$^{-1}$ is even closer to the experimental results from the main text. Performing 2LRT measurements on an $\approx 800$-nm-thick membrane with a heating laser wavelength of $\lambda_{\text{heat}}\,=\,325\,\text{nm}$ means that the corresponding light penetration depth (74\,nm) is smaller than the SM thickness $t$. As a result, not only in-plane thermal transport occurs, but also its cross-plane counterpart, as sketched in S-Fig.\,\ref{fig:surfacevolumeheating}(a). Thus, the measurement of pure $\kappa_{\text{in-plane}}$ values by 2LRT can only be guaranteed when the light penetration depth of the heating laser exceeds $t$ as sketched in S-Fig.\,\ref{fig:surfacevolumeheating}(b). With rising light penetration depth the SM is increasingly homogeneously heated in the cross-plane direction, which renders cross-plane thermal transport negligible. Nevertheless, we still expect our measured $\kappa$ values to provide a good estimate of $\kappa_{\text{in-plane}}$, because phonon transport in the cross-plane direction is hindered by the SM boundaries. For future measurements it would be promising to study even thinner SMs or to vary the penetration depth of the heating laser.\\

\newpage

One can observe in S-Fig.\,\ref{fig:kappa_accum_mfp} that phonons with $l_{\text{mfp}}$ values from $\approx 50$ to 300\,nm contribute most significantly to the final total value of $\kappa$ (in-plane or isotropic average). Please note that all graphs shown in S-Fig.\,\ref{fig:kappa_accum_mfp} converge to this final total value of $\kappa$ for $l_{\text{mfp}}\,>\,1\,\mu\text{m}$. While the presence of the phonon-boundary scattering will lead to an overall reduction in the $l_{\text{mfp}}$ values of all phonon modes, the phonon modes with a vanishing cross-plane velocity component will only be weakly affected. On the other hand, $l_{\text{mfp}}$ values will be strongly reduced for phonon modes with a large cross-plane velocity component. As such, the $\kappa_\text{in-plane}$ accumulation trend from S-Fig.\,\ref{fig:kappa_accum_mfp} has a similar profile for both SM thicknesses $t$ considered here. Further details regarding our theoretical treatment are given in the main text. Corresponding phonon dispersion relations and energy-resolved plots of the phonon scattering rates and $l_{\text{mfp}}$ values can be found in Ref.\,\onlinecite{elhajhasan_optical_2023} for a GaN SM with $t\,=\,250\,\text{nm}$.

\bibliography{AA)GaN_membrane1}